# Measurement of volume changes and associated stresses in Ge electrodes due to Na/Na$^+$ redox reactions

Subhajit Rakshit,[2] Akshay S. Pakhare,[1] Olivia Ruiz,[3] M. Reza Khoshi,[4]

Eric Detsi,[3] Huixin He,[4] Vijay A. Sethuraman,[5,6] Siva P.V. Nadimpalli[1,2,*]

[1]Department of Mechanical Engineering,

Michigan State University, East Lansing, Michigan, 48824, USA

[2]Department of Mechanical and Industrial Engineering,

New Jersey Institute of Technology, Newark, New Jersey 07102, USA

[3]Department of Materials Science and Engineering,

University of Pennsylvania, Philadelphia, Pennsylvania, 19104, USA

[4]Department of Chemistry,

Rutgers University, Newark, New Jersey, 07102, USA

[5]Department of Chemical Engineering,

University of South Carolina, Columbia, South Carolina, 29208, USA

[6]Faraday Laboratory LLC, Columbia, South Carolina, 29201, USA

[*]Corresponding author: E-mail: sivan@msu.edu

Tel: +1 517 432 2976, Fax: +1 517 353 1750

**Abstract**

*In situ* electrochemical cells were assembled with an amorphous germanium (a-Ge) film as working electrode and sodium foil as reference and counter electrode. The stresses generated in a-Ge electrodes due to electrochemical reaction with sodium were measured in real-time during the galvanostatic cycling. A specially designed patterned a-Ge electrode was cycled against sodium and the corresponding volume changes were measured using an AFM; it was observed that sodiation/desodiation of a-Ge results in more than 300% volume change, consistent with literature. The potential and stress response showed that the a-Ge film undergoes irreversible changes during the first sodiation process, but the subsequent desodiation/sodiation cycles are reversible. The stress response of the film reached steady-state after the initial sodiation and is qualitatively similar to the response of Ge during lithiation, *i.e.*, initial linear elastic response followed by extensive plastic deformation of the film to accommodate large volume changes. However, despite being bigger ion, sodiation of Ge generated lower stress levels compared to lithiation. Consequently, the mechanical dissipation losses associated with plastic deformation are lower during sodiation process than it is for lithiation.

1. Introduction

The recent push towards environmentally friendly energy production and carbon-free transportation technologies has renewed the interest in developing advanced energy-storage devices such as rechargeable batteries. Owing to an unparalleled volumetric and gravimetric energy densities among the available battery chemistries,[1,2] lithium-ion batteries (LIB) are the primary choice as energy storage device in portable electronics, electric vehicle, and grid-storage applications. However, projected widespread use of electric vehicles soon will increase the demand for lithium and drive the cost higher. Moreover, lithium reserves are limited and located in geographically-remote and conflict-affected areas of the earth. As a result, efforts to develop viable and alternative batteries such as aluminum-, magnesium-, and sodium-ion batteries have increased in recent years. Among these options, sodium-ion batteries (NIBs) are gaining momentum to be the potential alternatives for lithium-ion batteries, especially for grid-storage applications where low cost is the primary requirement. Sodium-ion batteries are cheaper due to the abundance of Na in the Earth's crust.[3–5] Further, Na does not react with Al[6], which enables replacing expensive Cu as the current collector, another practical advantage that makes sodium-ion batteries significantly cheaper.

Since Na is chemically similar to Li in many aspects, and the fundamental principles of NIB and LIB are identical, much effort has been dedicated to identifying electrode materials that are structurally similar to those used in lithium-ion battery technology.[5,7] Significant success has been achieved with such an approach in finding positive electrode materials for sodium-ion battery; for example, layered-transition metal oxides and tunnel-structured manganese oxides have shown to reversibly intercalate/de-intercalate Na-ions successfully, resulting in stable capacities of more than 140 mAh/g for several hundred cycles.[1] However, a search for suitable anode material is still in progress. Graphite, widely used negative electrode in lithium-ion batteries, is not suitable for sodium-ion batteries as it inhibits the intercalation of Na ions.[8] On the other hand, hard carbons have shown to react reversibly with Na producing a capacity of 300 mAh/g.[9] Other promising negative electrode materials for sodium-ion batteries include amorphous Ge (369 mAh/g),[10,11] Pb (485 mAh/g),[11,12] Sb (660 mAh/g),[11,13] and Sn (847 mAh/g).[13,14] Although these materials have comparable specific gravimetric capacities to those of Li-ion battery negative electrodes, they suffer from poor cyclic performance.[15,16] Among the available negative electrode material choices, Ge showed reasonable capacity retention; for example, Abel et al.[17] have showed that a nanocolumnar Ge retained 88% of the initial capacity for more than 100 cycles.

It has been shown in lithium-ion battery literature that volume expansion induced stresses dictate the long-term cyclic performance. For example, Si, Sn, and Ge expand almost 300% upon reacting with Li which induces a significant amount of stresses in these electrodes.[18–20] Sethuraman et al.,[21] Bucci et al.,[22] Al-Obedi et al.,[23] Nadimpalli et al.,[24] Pharr et al.,[25] and Soni et al.,[26] have experimentally showed that the magnitude of stresses in various electrode materials due to lithiation/delithiation cycling could reach as high as 1.5 GPa. These volume expansion induced stresses have been shown to cause extensive plastic deformation and fracture of electrodes resulting in rapid capacity fade.[27–30] It is also observed that the mechanical properties such as tensile modulus, Poisson's ratio, and yield stress vary with Li concentration,[31,32] this continuous variation of properties throughout battery operation will affect its cyclic performance. Besides being the driving force for mechanical damage and capacity fade, stresses also affect the equilibrium potential,[21] reaction kinetics,[33,34] and transport processes.[35] It is expected that the volume expansion induced stresses in sodium-ion battery electrodes will play a similar role and affect the cyclic performance of sodium-ion batteries. Hence, quantifying stresses generated due to sodiation/desodiation reactions is important to understand the damage evolution in sodium-ion battery electrodes. This information is necessary for designing damage tolerant and high-performance electrode architectures for NIBs. Significant amount of work has been done on the electrochemical behavior of various sodium-ion battery electrodes, but, despite its importance, their mechanical behavior has not yet been characterized. The lack of experimental data on the mechanical behavior also hinders the development of physics-based mathematical models for sodium-ion batteries.

Hence, the primary objective of this study is to measure the amount of volume expansion and the associated stresses in thin-film Ge electrodes during sodiation/desodiation reactions. To this end, sputter deposited Ge films (working electrode) on a double-side polished (DSP) fused silica wafers were cycled electrochemically against Na foil (counter/reference electrode) in a beaker cell. While the Ge film was cycled under galvanostatic conditions, the substrate curvature of DSP silica wafer was monitored using an optical technique to provide real-time stress measurements in the Ge thin film electrode. The volume change of a-Ge film due to sodiation/desodiation was determined by measuring the thickness of multiple Ge thin-film electrode samples that were sodiated/desodiated to different states of charge (SOC). The film thickness changed irreversibly after the first cycle, *i.e.*, the Ge film did not return to its original thickness after a full sodiation and desodiation cycle. It was observed that the steady state stress-capacity response of Ge film during sodiation/desodiation showed qualitatively similar behavior to that of a lithiated Ge film, but the magnitude of stress differs significantly. Despite a significantly bigger (Na-ion) size, the stresses generated due to sodiation are lower than the lithiation induced stresses in Ge thin film electrodes. The reported volume expansion data and real-time stress measurements of sodiated Ge will help develop mechanics-based models of sodium-ion battery electrodes and damage tolerant electrode design efforts.

## 2. Experimental methods

### 2.1 Germanium thin film electrode fabrication

Germanium thin films of 100 nm thickness were deposited on a fused silica substrate (thickness ~500 μm, diameter ~5.08 cm, double side polished) coated with thin films of Ti (~5 nm, as an adhesion layer) and Cu (~ 200 nm, as a current collector). The films were deposited by DC sputtering technique using Denton Explorer 14, and the inset in Fig. 1a shows the schematic of the thin-film configuration with various layers and their thickness values. To ensure uniform film thickness, the platen (which holds the samples) in the chamber was rotated at 20 rpm during the deposition process. After the deposition process, the thickness of the Ge film in all the samples was measured with a stylus profilometer (KLA-Tencor, KLA Corporation). The fused silica substrate ($SiO_2$) does not participate in any electrochemical reactions; it serves only as an elastic substrate for curvature measurements as shown in Fig. 1b. Similarly, the Cu current collector is inert to Na as per Na-Cu equilibrium phase diagram[36–38], and it will not contribute to the stress development during sodiation/desodiation cycling of Ge electrode. Sputter-deposited Ge thin films (under the above-mentioned conditions) are, in general, amorphous.[39] The XRD spectrum of an as deposited sample, shown in Fig. 1c, obtained by Bruker D8 Discover X-Ray Diffractometer (Bruker Corporation) confirms the amorphous nature of fabricated Ge thin films. The peaks shown in the XRD pattern, an indication of crystallinity, belongs to Cu, and absence of Ge peaks confirm its amorphous nature. In an earlier report, Nadimpalli *et al.*[24] confirmed the amorphous nature of these sputter deposited Ge films using Raman spectroscopy.

**2.2 Electrochemical cell assembly and measurements**

Fig. 1a shows the schematic of an in-house electrochemical cell, made of Teflon. The sodium-ion half cells were assembled and cycled at room temperature inside an argon-filled glove box (MBraun Inc., < 0.1 ppm $O_2$, < 0.1 ppm $H_2O$). The amorphous germanium (a-Ge) film as a working electrode; 1.3 mm thick sodium foil (prepared from Na cubes 99% trace metals basis, Sigma Aldrich) as a counter/reference electrode; and 1M sodium perchlorate ($NaClO_4$, >98% pure, Sigma Aldrich) in propylene carbonate (PC, 99.7% anhydrous, Sigma Aldrich) with 5 wt.% fluoroethylene carbonate additive (FEC, Sigma Aldrich) as an electrolyte were used to make electrochemical cells as shown in Fig. 1a. A glass microfiber sheet (pore size ~1 µm, Sigma Aldrich) was used as separator, which prevents any physical contact between electrodes, *i.e.*, avoids short circuit.

The amorphous Ge films were sodiated and desodiated under galvanostatic (*i.e.*, a constant current density of $i = 1$ µA/cm$^2$) conditions between 2 V and 0.001 V *vs.* Na/Na$^+$ using a Solartron 1470E potentiostat; the stress and potential response of the Ge film was recorded simultaneously during this process. Additional galvanostatic experiments with same current densities were performed on 30 nm and 120 nm thick Ge films to see if the electrochemical and stress response depends on film thickness. The cyclic voltammetry experiments were performed on the 100nm Ge films at a scan rate 0.1 mV/s between 2 V and 0.001 V vs. Na/Na$^+$. All the experiments reported here are carried out inside a glovebox (MBraun Inc, filled with Argon, <0.1 ppm O$_2$ and <0.1 ppm H$_2$O). Scanning electron microscopic analysis was carried out on as prepared and cycled samples. The cycled cells were dissembled, and the samples were rinsed with propylene carbonate for 10 min followed by 24 h of drying inside the glove box before removing them out for SEM analysis. The samples were carried in a sealed argon-filled container and transferred into the SEM chamber with minimum exposure to ambient air.

**2.3 Stress measurements using multi beam optical (MOS) setup**

Fig. 1b shows the schematic of the multi-beam optical sensor (MOS) set up (k-Space Associates, Dexter, MI), which was used to monitor the curvature evolution of fused silica substrate during sodiation/desodiation cycling of Ge film. The MOS set up consists of a solid-state laser source (with a central wavelength of 660 nm), a single collimated beam, and two etalons arranged to generate a 2x2 array of laser beams. The array of beams reflected from the substrate were captured by a CCD camera as shown in the schematic. The captured image appears as a 2x2 array of circular dots (cross-section of the laser beams) on a computer monitor. Sample curvature was determined by measuring the relative displacement of the laser dots as,

$$\kappa = \frac{\cos\emptyset}{2L}\left\{\frac{D-D_o}{D_o}\right\} \quad [1]$$

where $D$ is the distance between the center of the laser spots, $D_o$ the initial distance between the laser dots, $\emptyset$ the angle of the beams, and L the optical path length as shown in Figure 1b. The factor cos $\emptyset/2L$ is known as the mirror constant, which is specific to a given setup, was obtained by calibrating the setup using a mirror of known curvature. The 2x2 array of the laser spots enables curvature measurement in two orthogonal directions. This will be important to understand if the film expands isotropically and uniformly. Note, from Fig. 1a, that although the Na foil, glass fiber separator, and Ge films remain immersed in the electrolyte during the experiment, the surface of the silica substrate, which reflects the laser beams was not immersed. This was done to prevent optical complexities associated with laser traversing through the electrolyte.

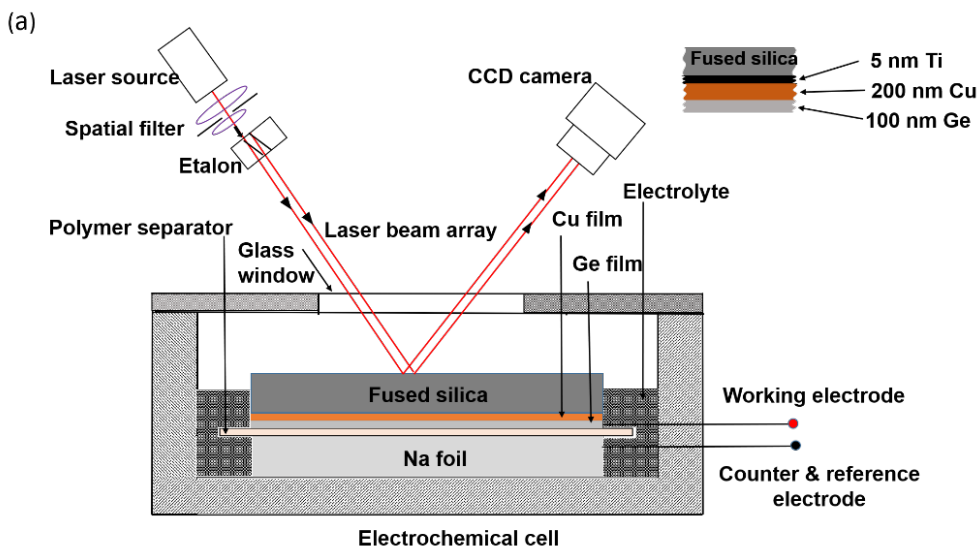

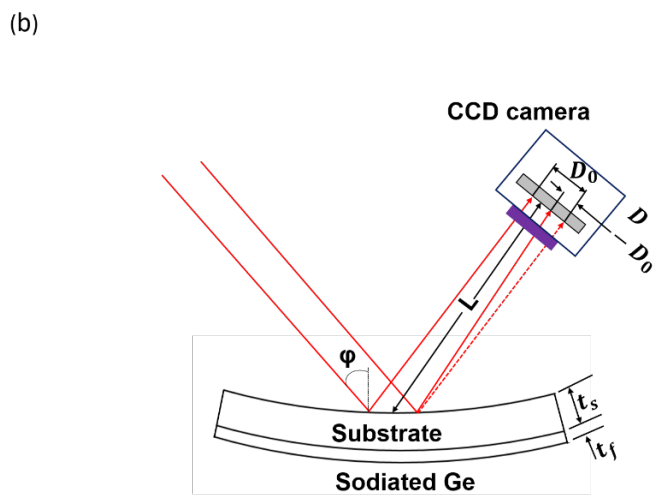

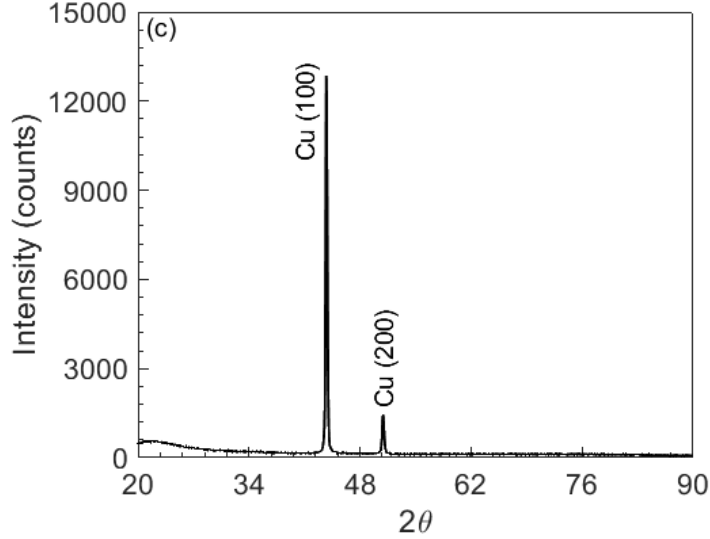

**Figure 1**. Schematic of (a) multibeam optical sensor (MOS) setup integrated with a custom-made electrochemical cell for enabling *in situ* curvature measurement. The inset shows the details of electrode samples, *i.e.*, various thin films and their thickness values, (b) shows details of MOS curvature measurement principle, and (c) depicts an XRD pattern of the as deposited films on silica substrate conforming the amorphous nature of Ge film (*i.e.*, no peaks for crystalline Ge).

The stresses in the Ge film are related to the substrate curvature by Stoney's equation,[31,40]

$$\sigma = \sigma_r + \frac{E_s t_s^2 k}{6 t_f (1-v_s)}, \quad [2]$$

where $E_s$, $t_s$, $v_s$ are the Young's modulus, thickness, and Poisson's ratio of the silica substrate, respectively. The parameter $\sigma_r$ is residual stress in the as prepared Ge film (caused by the thermal-expansion coefficient mismatch between the silica substrate and the Ge film as the film returns to room temperature from and elevated temperature during to deposition process), and $t_f$ is the thickness of the a-Ge film electrode, which changes continuously during sodiation/desodiation process. Based on the previous observations of Li-ion electrodes,[31,41–43] it is reasonable to assume that the volume expansion of Ge will be a linear function of sodium concentration, *i.e.*, the thickness evolution of Ge film as a function of capacity due to sodiation/desodiation is given as

$$t_f = t_f^0(1 + \beta z) \qquad [3]$$

Here, $t_f^0$ is initial film thickness, z is state of charge (SOC) which changes between 0 and 1; z =1 corresponds to a fully sodiated state with a theoretical capacity of 369 mAh/g and a volumetric strain of $\beta$.[11,44] The volumetric strain in Ge due to sodium reaction has not been measured experimentally before. Note from Eq.2 that the instantaneous film thickness $t_f$ is required to determine the true stress in the film. Hence, the thickness evolution of Ge film during to sodiation/desodiation reaction was measured here, the details of which are presented below.

**2.4 Volume change measurement of Ge due to sodiation/desodiation reaction**

A patterned Ge film as shown in Fig. 2a was fabricated by a sequence of nano- and micro-fabrication processes such as photolithography, e-beam evaporation, and lift-off (microposit remover) on a polished fused silica wafer (thickness ~500 μm, length = 25 mm, width = 25 mm). The thicknesses of Ge, Ni, and Ti films were 30 nm, 75 nm, and 5 nm, respectively; Fig. 2b shows the step-profile of a patterned film measured with an atomic force microscope (Dimension Icon, Bruker Corporation). The patterned Ge samples were then assembled as per Fig.1a in a half cell configuration and cycled under galvanostatic conditions to various levels of state of charge, and the changes in film thicknesses were measured. It should be noted that, due to constraint imposed by substrate, the area of the film is almost constant; hence, the volume change of Ge due to sodiation/desodiation reaction is reflected in thickness change.

Exposed silica substrate (shown in Fig. 2a), acts as a reference surface to measure the thickness expansion/contraction of Ge thin film electrode during sodiation/desodiation reaction; note that the thickness of Ni and Ti layers do not change as these layers do not react with sodium. Multiple samples were interrupted at various stages of cycling to obtain the volume change data; specifically, three samples were interrupted after the completion of first sodiation process, three samples were interrupted after one complete sodiation/desodiation cycle (*i.e.*, first cycle), and three more samples were interrupted after the completion of second sodiation (*i.e.*, sodiation process after $1^{st}$ cycle) . The interrupted cells were then dissembled, and the electrode samples were rinsed in propylene carbonate (PC) for 10 minutes followed by 24 h drying in the original glovebox (MBraun Inc.) before transferring them into a different glove box (MBraun Inc, filled with Argon, <0.1 ppm $O_2$ and <0.1 ppm $H_2O$) equipped with Bruker Dimension ICON AFM. The thickness measurements were carried out using the Peakforce tapping mode of scanning available in the AFM with a SCANASYST-AIR (spring constant 0.4 N/m) probe. In each sample, the thickness measurements were performed at multiple locations, and an average of all these measurements was taken as the representative thickness measurement from that sample. Fig. 2b shows the typical thickness data of as prepared (*i.e.*, uncycled) sample.

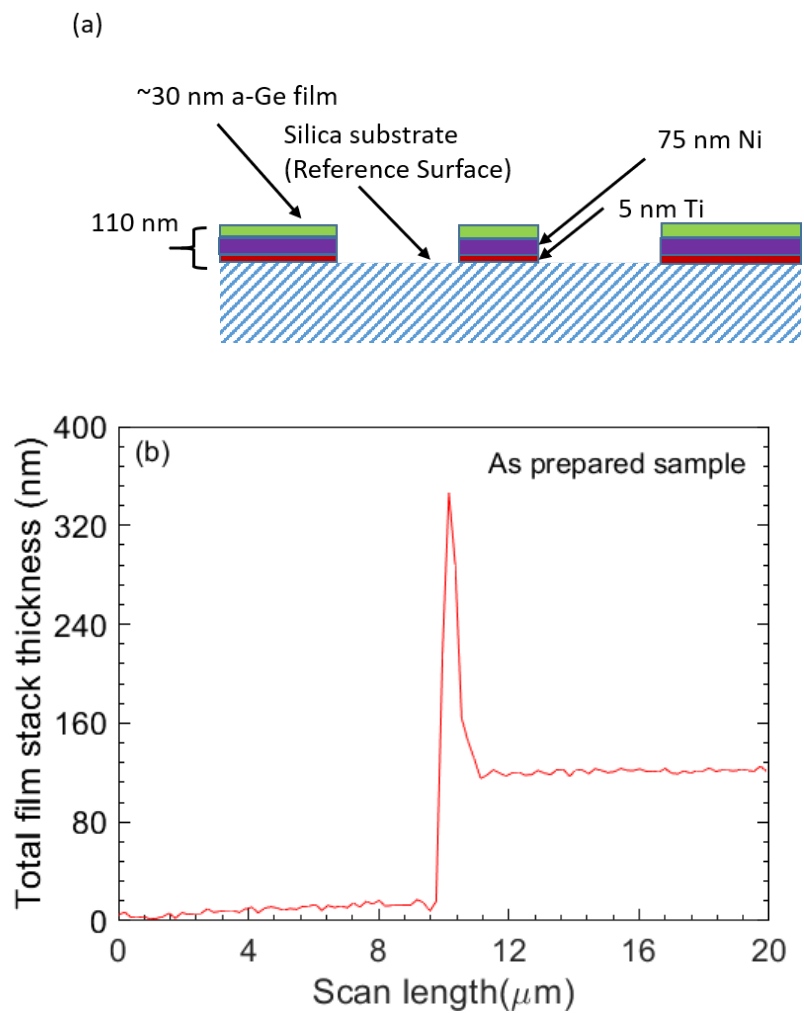

**Figure 2**. (a) Schematic of the patterned electrode sample used for volume expansion measurements; (b) the line profile of patterned sample shows the 110 nm stack of films (~5 nm Ti, ~75 nm Ni, and ~30 nm Ge). The measurements were performed at multiple locations in the sample to confirm the film thickness and step heights. The observed spike at the edge of film could be due to edge imperfections of lift-off process which was further amplified by the response of closed loop control system of the AFM.

## 3. Results & Discussion

### 3.1 Potential response of Ge electrode during sodiation/desodiation cycling

Fig. 3(a) shows the cyclic voltammogram of the thin-film a-Ge electrode cycled against a Na reference electrode at a scan rate of 0.1 mV/s in 1M sodium perchlorate in propylene carbonate with 5 wt.% fluoroethylene carbonate electrolyte. The cathodic sweep of cycle 1 corresponding to the sodiation of pristine a-Ge film shows a small peak (with a maximum at ~4.5 µA/cm$^2$) centered around 0.5 V *vs.* Na/Na$^+$, but absent in all subsequent cycles, is indicative of a passivation-type reaction corresponding to the formation of sodium- and chlorine-containing species due to solvent reduction. This is attributed to the formation of solid-electrolyte interphase (SEI) associated with the irreversible decomposition of the electrolyte species during first cycle. With continued scanning, the sodiation current sharply increases at 50 mV *vs.* Na/Na$^+$, and for subsequent sodiation cycles, the onset of this sharp increase in sodiation current happens at ~100 mV, suggesting improved kinetics with cycling. Note that in the context of a half-cell couple such as Ge(working)|Na(reference/counter), during the sodiation cycle, work is done by the cell on the potentiostat and therefore, higher potential *vs.* Na/Na$^+$ indicates facile kinetics. The reason for sluggish kinetics in the first sodiation of Ge could be attributed to additional kinetic barriers involved in the first cycle such as Ge-Ge bond breaking process. The difference in overpotential between the first desodiation cycle and subsequent desodiation cycles is marginal; note that during desodiation, work is done by the potentiostat on the cell and therefore, lower potential *vs.* Na/Na$^+$ indicates facile kinetics. Fig. 3(a) also shows that the desodiation curves for cycles 2-4 overlap indicating the same capacity and therefore good reversibility after first cycle.

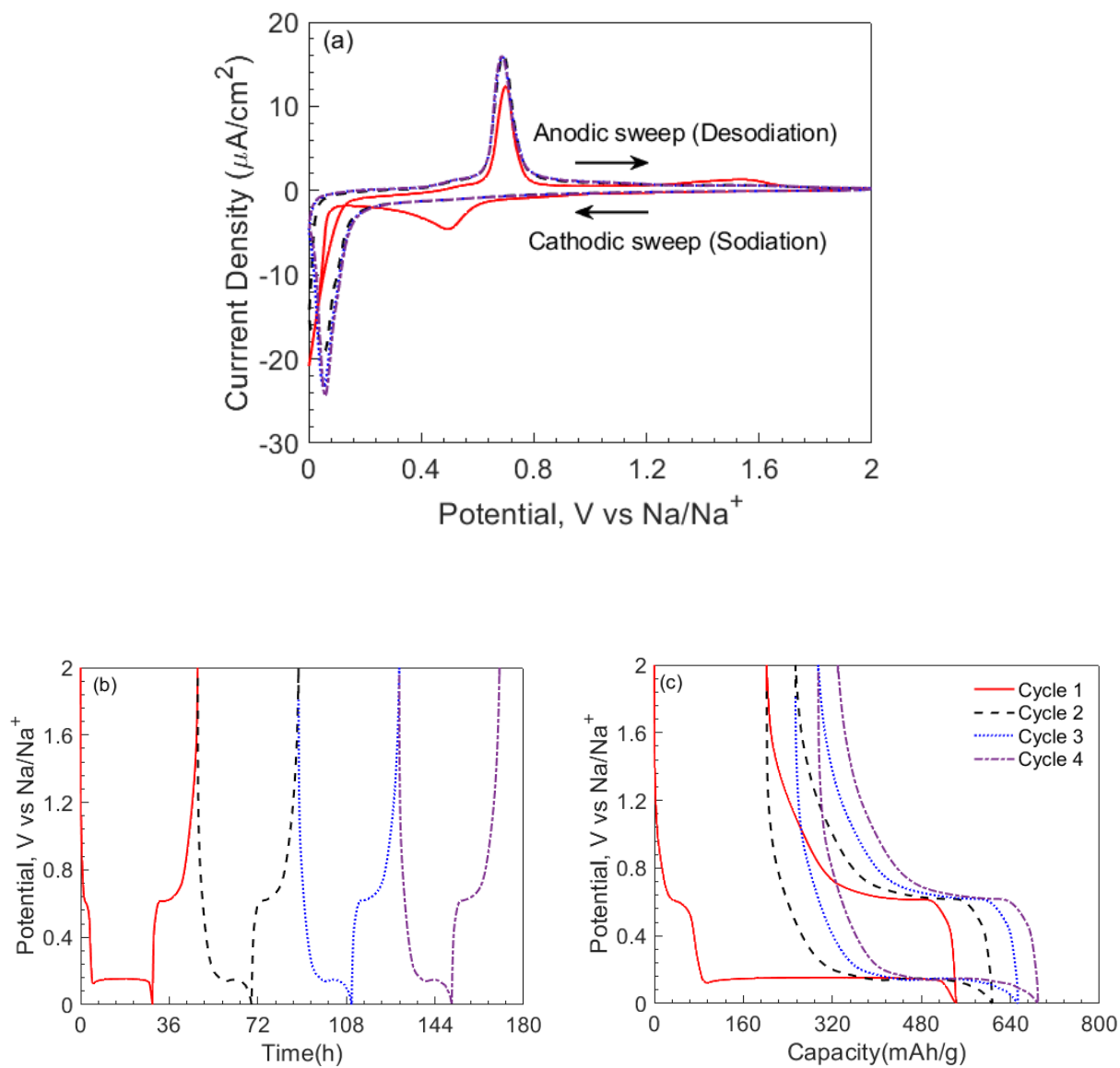

**Figure 3.** (a) Cyclic voltammogram of an a-Ge film at 0.1 mV/s scan rate against Na/Na$^+$. Potential response of Ge thin-film anode as a function of (b) time and (c) capacity during sodiation/desodiation cycling under galvanostatic conditions, i.e., at a constant current density of ~1 μA/cm$^2$ (*ca.* C/20 rate).

Figs. 3b and 3c show the potential response of the Ge film as a function of time and specific capacity, respectively, during sodiation/desodiation cycling under a constant current density of 1 μA/cm$^2$ (which corresponds to ~C/20 rate). At the beginning of the first cycle, the potential of the cell drops sharply from an open-circuit value of 2 to 0.61 V *vs.* Na/Na$^+$ where it remains constant for a short period of time before decreasing sharply to a potential plateau of 0.14 V *vs.* Na/Na$^+$, which corresponds to sodium alloying with the a-Ge film. The potential remains almost constant at 0.14 V *vs.* Na/Na$^+$ until the end of sodiation and drops rapidly to the cut off value of 0.001 V *vs.* Na/Na$^+$ at 540 mAh/g. Upon desodiation, the potential rises sharply to 0.65 V *vs.* Na/Na$^+$ within a 40 mAh/g change in capacity and remains constant at 0.65 V *vs.* Na/Na$^+$ until the capacity decreases to 360 mAh/g, corresponding to sodium extraction from Ge film, and gradually rises to 2 V *vs.* Na/Na$^+$ at the end of desodiation process, *i.e.*, at 200 mAh/g. The columbic efficiency in the first cycle was 63%, but it increased to 78% in the second and to 81% in the fourth cycle. The short plateau of 0.61 V *vs.* Na/Na$^+$ that is present in the first cycle sodiation process but almost nonexistent in the rest of the cycles, Figs. 3b and 3c, can be attributed to irreversible side reactions, which lead to the formation of the SEI layer. These observations are in good agreement with those reported by Baggetto *et al.*[11]

The flat potential response such as 0.14 V *vs.* Na/Na$^+$ during sodiation of fresh a-Ge film in Figs.3b and 3c is, in general, an indication of a two-phase reaction in the electrodes, *i.e.*, a reaction which leads to formation of a sharp phase boundary which separates two equilibrium phases with a sharp concentration jump across the phase boundary. This sharp phase boundary then propagates into the film until the entire film is consumed by a single phase. Some examples of such phenomenon are crystalline to crystalline phase transformation when Li reacts with graphite *via* stages, Al, and Sn to form Li$_x$C,[45] Li$_x$Al intermetallics[46] and, Li$_x$Sn intermetallics [15,47] respectively; and crystalline to amorphous phase transformation when Li reacts with crystalline Si to form amorphous Li$_x$Si. In fact, it was reported [14,48,49] that sodiation of Sn also exhibits flat potentials indicating two-phase reactions. However, one significant difference between these previous reports and the present study is that unlike the above reports where the starting samples are crystalline in nature at the beginning of the experiment, the Ge thin film electrodes in the present study are amorphous in nature to start with, see Fig. 1c. In addition, as per the *in situ* X-ray diffraction measurement of Baggetto *et al.*[11], the a-Ge film remains amorphous during and at the end of sodiation, suggesting the formation of a Na$_x$Ge solid solution, yet a flat potential response, a feature of two-phase reaction, was observed in the current study and the earlier report of Baggetto *et al.*[11]. A thorough and extensive *in situ* characterization (*i.e.*, TEM and XRD) needs to be carried out to shed more light on this behavior, which is beyond the scope of this study.

It should be noted, however, that a flat potential need not always be associated with a phase transformation. For example, the *in situ* TEM work on a-Si by Wang *et al.*[50] and McDowell *et al.*[48] indicated that under extremely fast lithiation/delithiation rates or diffusion limited conditions, appearance of two-phase reaction in amorphous electrode materials (or a-Si) during electrochemical (lithiation) reaction can occur. McDowell *et al.*[48] observed this seemingly two-phase type reaction (a sharp phase boundary moving through a-Si particle) occurring only during the first lithiation process but not in the subsequent cycles; the reaction in subsequent cycling occurred via a single phase mechanism (formation of solid solution indicating a higher sloping voltage behavior).[51,52] McDowell *et al.* argued that, this behavior which is similar to the observations of this study (Fig. 3), was due to the rate-limiting effect of Si-Si bond breaking relative to the lithiation rate. It should be noted that a-Si does not exhibit this behavior under very slow lithiation rates as shown by Sethuraman *et al.*,[31] Bucci *et al.*,[22] Nadimpalli *et al.*,[30] and Pharr *et al.*[53] It is expected that a similar rate limiting phenomena, such as relatively low Na diffusion in pristine Ge compared to cycled Ge as suggested by CV curve in Fig. 3a, is responsible for a flat potential when a pristine a-Ge film is sodiated (resembling a two-phase reaction mechanism) and a single phase reaction in the subsequent cycles. In fact, this behavior was reflected in the stress measurements which are presented below.

## 3.2 Stress response of Ge thin film electrode during sodiation/desodiation reaction

Figures 4a and 4b show the stress-thickness (or curvature) evolution of a-Ge film as a function of time and capacity, respectively, during a galvanostatic sodiation/desodiation cycling. Note that the stress-thickness value at zero capacity is not zero because of residual stress. In general, thin films are deposited at elevated temperatures, and due to thermal expansion property mismatch, residual stresses develop in the films when the film/substrate system is cooled to room temperature. The residual stress in all the experiments reported here was obtained by measuring the substrate curvature (with setup shown in Fig. 1) immediately before and after depositing the films and using Eq. 2 (only the second term on the right-hand side). The average value of the measured residual stress in the films was -0.24 GPa; the sign represents the state of stress: negative for compressive and positive for tensile stress.

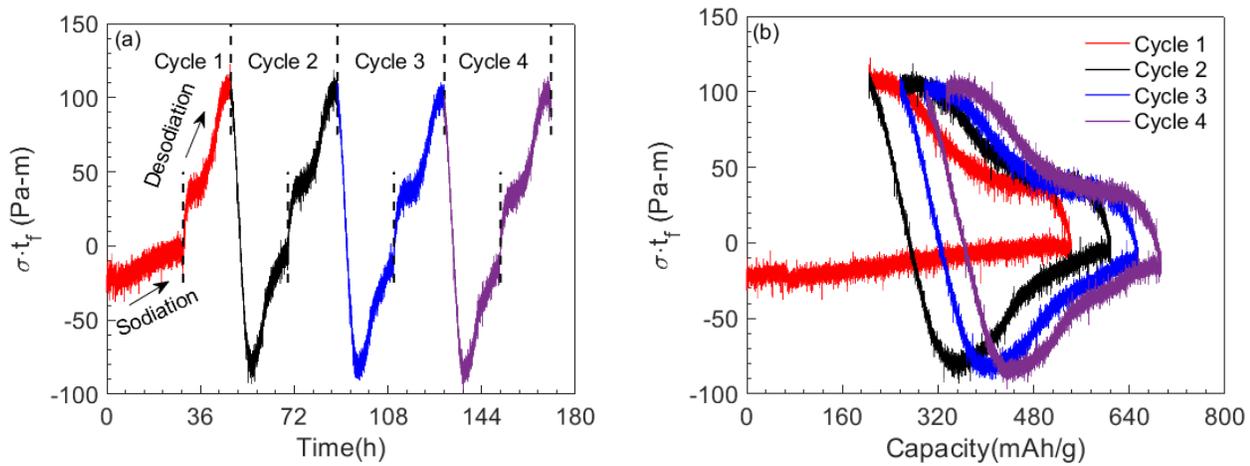

**Figure 4** Stress-thickness behavior of a-Ge thin film anode as a function of (a) time and (b) specific capacity, respectively, during sodiation/desodiation under galvanostatic cycling (at constant current density of ~1 μA/cm$^2$, i.e., at C/20 rate).

Fig. 4b shows that the stress-thickness value remained almost constant with capacity when sodium reacted with pristine a-Ge, reaching a value of – 2 Pa-m at the end of sodiation (543mAh/g). Upon desodiation, the stress-thickness value increased rapidly to 43 Pa-m within a 40 mAh/g change in capacity and thereafter increased gradually and nonlinearly with decreasing capacity, reached a peak value of 109 Pa-m at 204 mAh/g (at the end of desodiation). Although the desodiation response is similar in all the cycles, the response of the film during the first sodiation (i.e., when sodium reacted with pristine a-Ge film) is strikingly different from that observed in the subsequent cycles, i.e., in subsequent cycles, the stress-thickness value did not remain constant but increased rapidly (and linearly) with capacity to a peak value of -93 Pa-m (at 355 mAh/g in the second sodiation process) and decreased thereafter non-linearly with further sodiation, reaching a value of -8 Pa-m at the end of sodiation, mirroring the behavior observed during desodiation process. The in-plane expansion of Ge film during sodiation is constrained by substrate which induces biaxial compressive stresses in the film. Similarly, the substrate prevents the in-plane shrinking of the film during desodiation inducing biaxial tensile stresses; however, the film is free to expand in the normal (or thickness) direction.

It should be noted that the curvature values measured may lead to error in stress estimates if the film cracks during cycling. The fact that the peak stress-thickness value of ~114 Pa-m during desodiation and ~ -94 Pa-m during sodiation remained almost constant for the first four cycles (Fig. 4a) is an indication that the Ge film did not develop cracks during the first four cycles. The SEM analysis carried out on various samples, shown in Fig. 5, confirms this observation that the Ge film was intact for the first four cycles and developed cracks only during 5$^{th}$ cycle. Although the mechanism of crack initiation in amorphous Na$_x$Ge is not understood at present, a closer look at the surface feature of the electrode in Fig.5 along with some recent literature may offer some preliminary hypothesis. Note from Fig. 5 that sodiation/desodiation cycling leads to increased surface roughness, and it may also cause porosity[11,54]. Further, both the roughness and porosity evolve with cycling. The formation of cracks under evolving surface features, microstructure, and repeated tensile/compressive stress cycling in Na$_x$Ge film is similar to low cycle fatigue failure in polycrystalline structural materials. The stress response shown in Fig. 4 was repeatable in at least 10 different samples fabricated in different batches, and also the curvature response was independent of the in-plane direction of the film (see Fig. S1), which indicates that the expansion of the film due to sodiation/desodiation reaction is isotropic. Further Figs. S2-S4 show that the stress response of the film was also independent of film thickness (i.e., 30 nm, 100 nm, and 120 nm a-Ge thin films showed negligible variation in measured data). In other words, the response shown in Fig.4 is repeatable in several samples and is independent of film thickness (at least in the range from 30 nm to 120 nm).

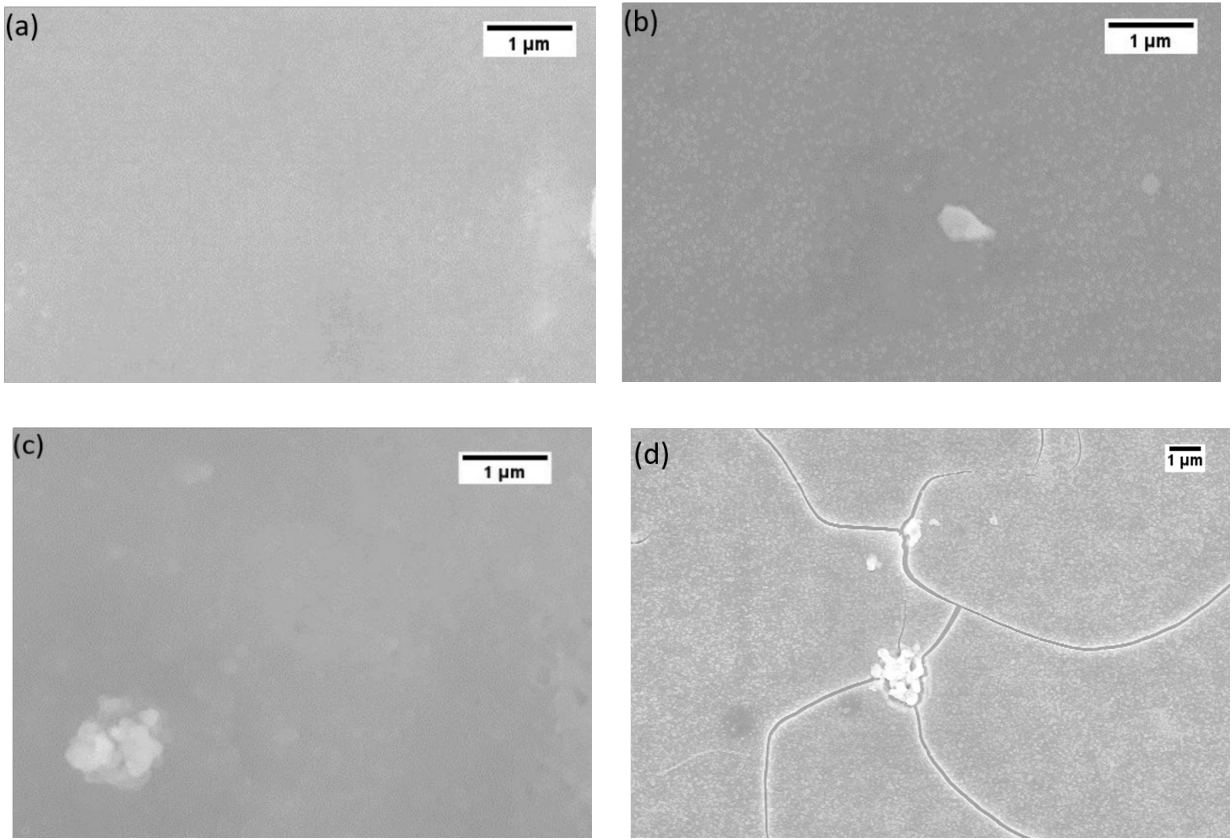

**Figure 5.** Surface morphology study of as deposited and cycled Ge thin film electrodes by SEM (scanning electron microscopy): (a) pristine a-Ge film, (b) Ge film after first sodiation, (c) Ge film after sodiation/desodiation cycle, (d) Ge film after 5$^{th}$ cycle.

The stress-thickness response of a-Ge, Fig.4, is qualitatively similar to that of crystalline Si during lithiation/delithiation under galvanostatice conditions. Chon *et al.* [55] reported that when crystalline Si first reacts with Li, it exhibits a two-phase reaction resulting in flat potential and linear stress-thickness response, which is strikingly different from the response in subsequent cycles. This behavior of Si is attributed to irreversible change of crystalline Si to amorphous $Li_xSi$ during the first lithiation process, which remains amorphous for the remaining cycling history.[56] The distinctly different stress response of a-Ge film during first sodiation process compared to the response in subsequent cycles shown in Fig.4 can be attributed to the irreversible changes that occur in the a-Ge film during the first sodiation process. The reason for almost flat stress response can be explained as follows. Under a diffusion limited reaction between sodium and pristine Ge (i.e., first sodiation), as suggested by flat potential response in Fig. 3, a sharp concentration front divides the Ge film into sodium-rich and sodium-poor regions, indicated as red and blue regions, respectively, in the schematic of Fig.6a. A flat potential further suggests that the concentration of sodium in the sodium-rich Ge remains constant while the front propagates into the remaining Ge film. During this process, both (blue and red regions in Fig. 6a) regions of the Ge film contribute to the measured stress as $\sigma t_f = \sigma_1 t_1 + \sigma_2 (t_f - t_1)$. Here $\sigma_1$ and $t_1$ are the stress and thickness values, respectively, of sodium-rich (or red) region, and $\sigma_2$ is the stress in the sodium-poor (or blue) region during initial sodiation. Note that $\sigma_1$ and $\sigma_2$ remain constant as the concentration of sodium in both regions remains constant. Further, it is reasonable to assume that $\sigma_2$ is the residual stress (i.e., stress in Ge film at the beginning of the experiment or at low sodium concentration, Figs. 4 and 6), and $\sigma_1$ to be less than $\sigma_2$ as the sodium-rich film stress is almost close to 0 GPa from Fig. 4. Hence, the measured stress during first sodiation is given as $\sigma t_f \cong \sigma_2 (t_f - t_1)$ which remains almost constant or decreases slightly as $t_1$ increases, resulting in a flat response shown in the first sodiation process in Fig. 4. Once the film completes this irreversible conversion in the first sodiation process, the response of the film in the remaining cycling history is repeatable, consistent with the potential response of Fig. 3.

Additional experiments were carried out to test the hypothesis that unless the entire pristine a-Ge film is activated/reacted (or converted irreversibly), the flat response (or marginal change in stress-thickness) with capacity, as seen in the first sodiation process of Fig. 4, will occur during sodiation in the subsequent cycles. Figure 6b and 6c show the potential and stress-thickness data, respectively, from one such experiment where the Ge film was partially sodiated (or activated) and desodiated, with each cycle sodiating deeper into the film, i.e., progressively converting (or activating) more, until the entire pristine a-Ge has irreversibly reacted. The current density $i = 3$ µA/cm$^2$ was constant throughout this experiment. Note that the stress-thickness value did not change during the first 4 h sodiation/ desodiation cycle because the majority of this sodiation was SEI reaction. However, in the next cycle when the film was sodiated for 5 h, as anticipated the curvature remained almost constant, and the desodiation response is similar to what was expected according to Fig.4, though the magnitude is different because only a small portion of the film has been activated in this cycle and contributed to this curvature change. During the third cycle sodiation process, the portion of the film that was activated earlier contributed to the initial sharp linear increase in stress-thickness with capacity, and when the sodium reacted with remaining pristine a-Ge, the stress-thickness value remained almost constant as per the hypothesis. Finally, when the entire film is activated (or irreversibly converted), the stress-thickness response during the desodiation and subsequent cycling matched both qualitatively and quantitatively with Fig.4. This confirmed the hypothesis that the initial reaction of Na with pristine a-Ge activates the film and converts it irreversibly.

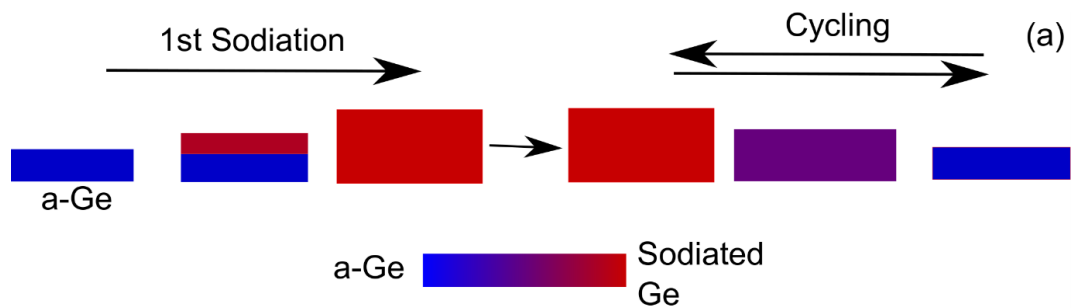

(a)

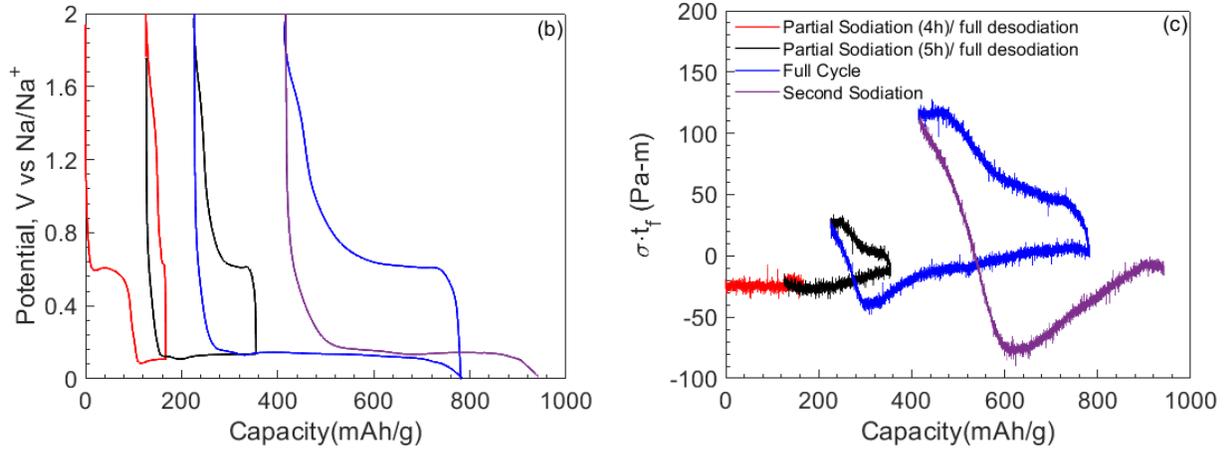

**Figure 6**. (a) Schematic of irreversible reaction mechanism during 1st sodiation reaction of pristine a-Ge thin film followed by reversible single phase sodiation/desodiation mechanism, as per the observations of Figs. 3 and 4. (b) and (c) show the potential and stress-thickness response of the a-Ge film as a function of capacity, respectively, during partial sodiation/desodiation cycling under galvanostatic conditions with each cycle progressively reacting more of the pristine a-Ge film. Note that the distinct signature of fresh germanium reaction with sodium, i.e., almost negligible curvature change with capacity, occurring in subsequent cycles until all the fresh Ge is reacted with Na.

3.2.1 *Volume changes and associated true stress due to sodation/desodiation cycling of Ge film*

The data presented in Figs.4 and 6 is directly obtained from the curvature measurements, and to determine the true stress (or Cauchy stress) in the film at any given sodium concentration, the instantaneous film thickness is required as per Eqs. 2 and 3. Figs. 7a and 7b show the AFM image of a patterned a-Ge electrode before and after the first sodiation process, respectively; and Fig. 7c shows the height profiles of Ge thin films sodiated/desodiated to different stages of cycling. The surface of (inert) silica substrate is used as reference for thickness measurements as it will not participate in electrochemical reactions. Note from Figs. 7b and 7c that the surface of a-Ge film develops noticeable roughness after reacting with sodium; for example, note from Fig. 7c that the thickness value of the sodiated film fluctuates more compared to that of desodiated film. As a result, the thickness value averaged over the scan length, indicated with a dashed-line, was considered as the representative thickness value of the sample. A minimum of three samples were considered for each SOC and a minimum of two locations in each sample were scanned. As mentioned earlier, the substrate constraint keeps the area of the film almost constant; consequently, the volume change is directly proportional to thickness change in the current measurements. Fig. 7c summarizes all these thickness (or volume) change measurements due to sodiation/desodiation cycling.

Figure 7d shows the ratio of sodiated Ge film's instantaneous thickness to its original value (i.e., pristine film thickness) as a function of sodiation/desodiation history. Each data point (open circle) corresponds to a value from a single sample and the filled triangles represent the average value of the ratio $\frac{t_f}{t_f^o}$ for different sodiation/desodiation states of the a-Ge film. Note that a 30 nm pristine a-Ge film expanded to 101 nm, corresponding to an average $\frac{t_f}{t_f^o} = 3.39$, at the end of full sodiation. Upon desodiation, i.e., at the end of first cycle, the film shrinks to 62 nm, but does not return to its original thickness of 30 nm. This is expected, because as per thermodynamics it is impossible to remove all the Na that reacted with a-Ge. This is also consistent with the observation of Figs. 3 and 4 that the pristine a-Ge film undergoes irreversible reaction during the first sodiation process. Upon the second sodiation process, the film expands to an average value of 4. This measured value of the expansion ratio of 3.39 and 4 for a fully sodiated state is slightly greater than the value of 3 estimated by Lu et al.[57] based on the expansion of sodiated Ge nano tube. It should be noted that thickness measurements in the current study did not include the effects of SEI layer formation and its thickness. Also, the volume change comparison is being made from different samples (interrupted at various stages of cycling) as opposed to *in situ* thickness measurement of a single film going through sodiation/desodiation cycling. Hence, various factors such as AFM scan parameters, sample to sample variation, non-uniform reaction across the sample, and nanopores formation during desodiation contribute to the scatter. Nonetheless, the data obtained here is, to the author's knowledge, one of the first attempts of measuring volume expansion of a-Ge due to reaction with sodium. Due to the various assumptions made in these volume expansion measurements, the measurements should be considered first order estimates. Although the thickness measurements were only made at the terminal points (i.e., at the beginning and end of sodiation/desodiation), as argued earlier, the thickness values at any intermediate SOC can be estimated by linear extrapolation according to Eq.3. Considering the expansion of Ge film from experimental measurements, the thickness expansion equation 3 can be re-written as

$$t_f = t_f^0(1 + 2.39z) \qquad [4]$$

where $t_f^0$ is initial film thickness, z is state of charge (SOC) which changes between 0 and 1; z =1 indicates to a capacity of 369 mAh/g and a volumetric strain of 2.39 obtained from experiments (i.e., Fig. 7).

The true stress in the a-Ge film due to sodiation is obtained by using the estimated a-Ge film thickness during sodiation/desodiation cycling as per Eq. 4 and the stress-thickness data presented in Fig. 4. The true stress data of a-Ge film during the stable 2$^{nd}$ cycle is plotted in Fig.8a. As noted earlier, upon sodiation the stress increases linearly (elastic response), reaches a peak compressive stress value of -0.56 GPa at 100 mAh/g, and decreases slowly with capacity reaching -0.18 GPa at the end of sodiation. The nonlinear stress response beyond 100 mAh/g capacity suggests that the sodiated Ge film undergoes extensive plastic deformation to accommodate the volume expansion. Upon desodiation the stress became tensile (i.e., positive value) and remained close to 0.01 GPa until the capacity decreased to 270 mAh/g and then increased to a peak value of 0.75 GPa at the end of desodiation. The stress response during desodiation almost mirrors the response during sodiation process. Since the Ge film continues to deform plastically, the measured stress history in Fig. 8a can be considered as the evolving yield strength of the Na$_x$Ge alloy as a function of sodium concentration. Although the stress response of a-Ge during sodiation/desodiation, i.e., in Fig. 8a is qualitatively similar to the response of a-Ge during lithiaton/delithiaton, the magnitude of the stresses differs significantly. A detailed comparison of stress response of Ge due to lithiation versus sodiation is presented below.

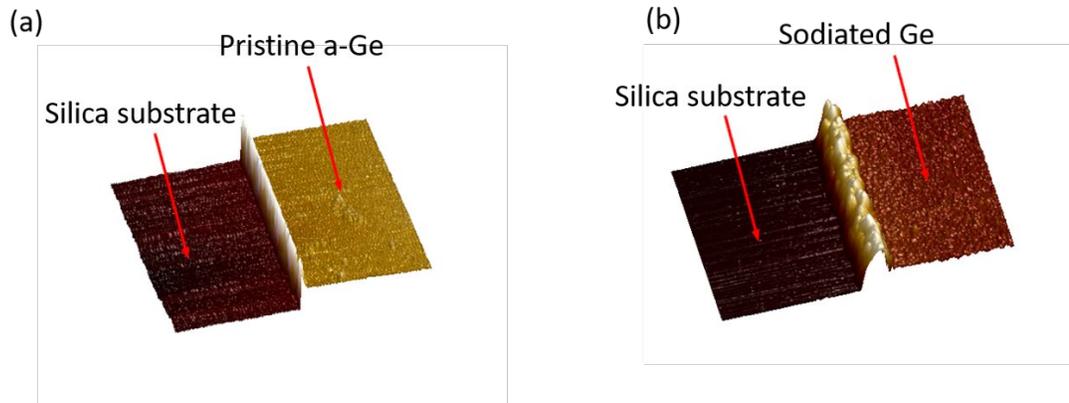

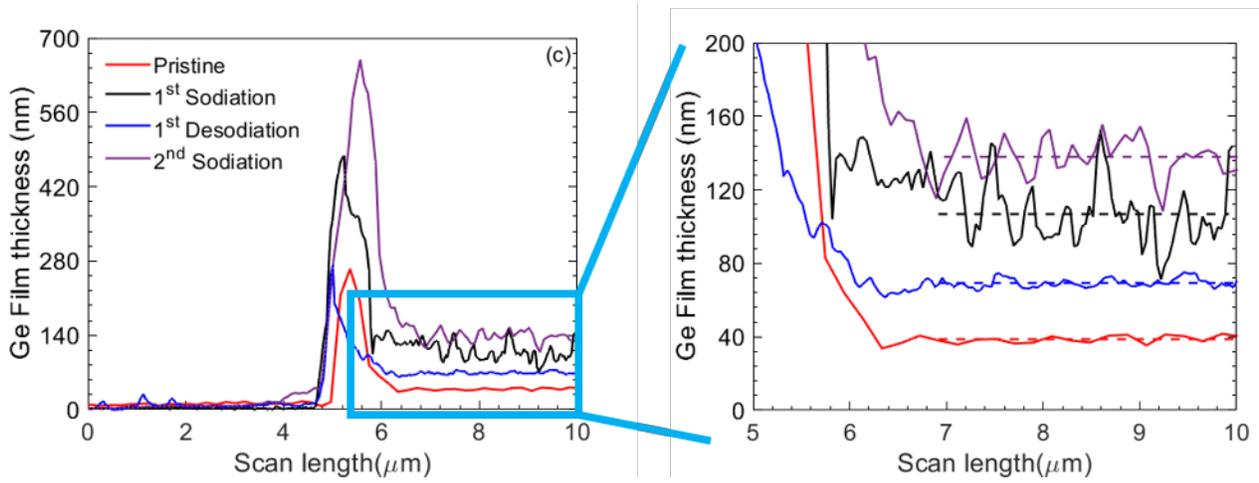

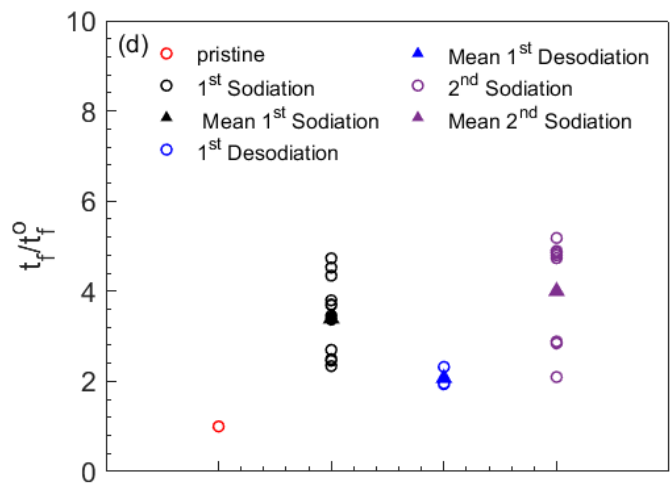

**Figure 7.** AFM image of (a) pristine a-Ge film and (b) the film at the end of 1st sodiation process; (c) thickness profile of a-Ge films sodiated/desodiated to different levels of state of charge; and (d) ratio of final to initial thickness values of a-Ge film at various stages of sodiation/desodiation cycling.

4. **Effect of ion size ($Na^+$ vs $Li^+$) on the mechanics and electrochemistry of Ge film**

Figs. 8a and 8b show the stress evolution in a-Ge film during galvanostatic sodiation/desodiation and lithiation/delithiation cycling, respectively. The current density during sodiation/desodiation cycling and lithiation/delithiation cycling of Ge was 1 µA/cm² and 5 µA/cm², respectively. Note that initially the stress response is linear with capacity during both sodiation and lithiation, but the peak compressive stress of sodiated Ge is slightly lower than that of lithiated Ge. After reaching peak the stress continues to decrease during sodiation, but during lithiation process the stress remains almost constant at -0.5 GPa until the end of lithiation process. As a result, the stress level in sodiated Ge is always less than that of lithiated Ge. It should be noted that the peak compressive stress due to lithiation could reach as high as -1.1 GPa [24] which is significantly higher than -0.56 GPa observed in sodiated Ge. This is interesting because the Li, in spite of being a smaller atom compared to Na, can induce a peak stress that is twice the magnitude than what Na can induce when reacting with Ge. Moreover, the stress at any concentration in $Li_xGe$ is higher (both during charge/discharge) than the stresses in $Na_xGe$. However, the bigger size of the sodium atom is one of the reasons why the volume change due to sodiation is more than 300%, comparable to that of lithiation induced volume change, even though only one Na atom per Ge is inserted into the network. In other words, Ge network can accommodate significantly more Li atoms than Na atoms for a similar volume change. For example, full lithiation of Ge results in the formation of $Li_{15}Ge_4$, i.e., 3.75 Li atoms per Ge atom, but the full sodiation results in NaGe, i.e., only one Na atom per Ge atom in the network[11,17,44]. Due to the low stress values at almost similar strain (volume change) values in sodiated Ge, as will be shown below, the mechanical dissipation losses associated with plastic deformation of electrodes is relatively low in sodiated Ge compared to that of lithiated Ge. It should be noted that lower stresses in electrodes during cycling may also reduce the mechanical damage, i.e., cracking.

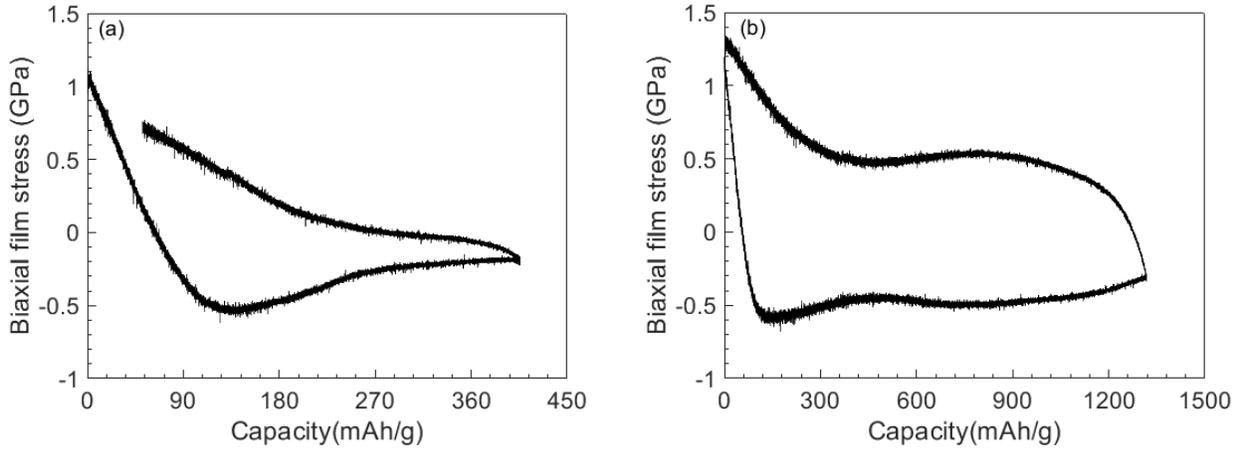

**Figure 8**. Stress response of a-Ge film as a function of specific capacity during (a) sodiation/desodiation and (b) lithiation/delithiation cycling under galvanostatic conditions obtained from Nadimpalli *et al.*[24]

As the efficiency of the battery is affected by electrode stresses and associated plastic deformation, it is instructive to quantify the losses associated with it. An energy balance-based procedure given by Sethuraman *et al.* [31] to estimate the losses due to plastic deformation in electrodes is adopted here to estimate the energy loss during sodiation/desodiation of Ge. The energy balance during sodiation can be written as,

$$W_\mu = W_c^s + W_m^s + W_p^s. \qquad [5]$$

Here, $W_\mu$ is the energy available in the half cell, $W_c^s$ is the work done by the cell on the potentiostat, $W_m^s$ is the mechanical dissipation in Ge electrode due to compressive plastic deformation, and $W_p^s$ is the sum of all other dissipations due to various polarizations (i.e., charge transfer, ohmic, and transport). The $W_c^s$ is calculated as

$$W_c^s = I \int_0^{t_s} V \, d\tau, \qquad [6]$$

where $V$ is cell potential, $I$ is the current, and $t_s$ is the total sodiation time. The $W_c^s$ is calculate to be 0.19 J. The energy loss due to mechanical dissipation $W_m^s$ is calculated as

$$W_m^s = 2A \frac{c_{max}}{t_s - t_p} \frac{d\epsilon}{dc} \int_{t_p}^{t_s} \sigma_f \, t_f \, d\tau, \qquad [7]$$

where $A$ is the total film area, $c_{max}$ is the stoichiometric ratio of Na to Ge at the end of sodiation, and $t_f$ is the film thickness. The $\frac{d\epsilon}{dc}$ was obtained from the measured volumetric strain data to be 0.50 and $c_{max}$ is taken to be as 1. The integration in Eq. 7 was carried out in the plastic regime (i.e., between $t_p$ the time at the start of plastic deformation and $t_s$ the time at the end of sodiation) of the stress-thickness curve. The calculated value of $W_m^s$ for sodiation of a-Ge is 0.04 J.

A similar energy balance for the desodiation process gives the work done by the potentiostat on the cell as

$$W_c^d = W_\mu + W_m^d + W_p^d. \qquad [8]$$

The quantities in this equation are same as before except the superscript $d$ which stands for desodiation. Using a similar procedure, the $W_c^d$ and $W_m^d$ are calculated to be 0.53 J and 0.07 J, respectively. Considering polarization losses ($W_p^s = W_p^d$) to be same during sodiation and desodiation and eliminating $W_\mu$ from both equations 5 and 8, the total energy loss due to the polarization in one complete cycle ($W_p^s + W_p^d$) is obtained as 0.23 J. A similar energy balance analysis of the lithiatied Ge data of Nadimpalli et al.,[24] shows that the polarization losses in lithiated Ge half-cell are approximately 0.19 J and mechanical dissipation loss due to lithiation and delithiation of Ge to be 0.34 J. Note that, in spite of significantly low capacity (369 mAh/g Na$_x$Ge vs. 1320 mAh/g of Li$_x$Ge), the polarization losses in one complete cycle are almost similar for both cases; this can be attributed to the sluggish kinetics of sodiation/desodiation reaction. However, the contribution of plastic deformation to total energy loss in a cycle is approximately 64 % in Li$_x$Ge compared to only about 32 % in Na$_x$Ge. Hence, despite a bigger ion size, sodiation of Ge results in lower electrode stresses and lower mechanical losses compared to lithiation.

## 5. Conclusions

The volume expansion and the associated stresses during galvanostatic sodiation/desodiation of amorphous Ge thin-film electrodes have been quantified. Both the cyclic voltammetry and galvanostatic measurements showed that the reaction kinetics are sluggish and the a-Ge film undergoes irreversible changes during the first sodiation process, but in the subsequent cycling process, not only the kinetics and columbic efficiency improve but also the film response becomes reversible. The galvanostatic sodiation of a-Ge shows a long flat potential of 0.14 V *vs*. Na/Na$^+$ during first sodiation, an indication of a possible two-phase reaction, which is not present in the subsequent cycling. These observations are in good agreement with the literature.

The stress-thickness response of a-Ge film during sodiation/desodiation cycling was measured using a multi-beam optical sensor (MOS) setup. It was observed that the response of the film during the first sodiation (i.e., when sodium reacted with pristine a-Ge film) is strikingly different from the response in subsequent cycles, which is consistent with the potential response of the film, i.e., the first sodiation of pristine a-Ge results in an irreversible response. The Ge film remained intact (i.e., no cracking and delmination) for the first five sodiation/disodiation cycles, and the response of the film due to sodiation/desodiation reaction is isotropic.

Volume changes of a-Ge film due to sodiation/desodiation reaction was measured using an AFM housed inside a glovebox with inert argon environment. It was noted that the surface of the electrode develops significant roughness after reacting with sodium. During sodiation process, the pristine a-Ge film experienced 239% volume expansion by the end of full sodiation. Upon desodiation, i.e., at the end of first cycle, the film shrinks, but does not return to its original thickness. This is consistent with the electrochemical and stress-thickness data. Upon the second sodiation process, the film undergoes an average volume change of 300%. These thickness measurements should be considered as first order estimates, because the influence of SEI formation was not taken in to account.

The stress induced in Ge film due to sodiation was obtained by using the measured curvature and thickness evolution data in the Stoney's equation. It was observed that the stress response becomes repeatable after first sodiation process. Initially, the stress (compressive) increases linearly with capacity during sodiation and reaches a peak value of -0.56 GPa at 100 mAh/g and there after the stress response becomes non-linear and film continues to deform plastically to accommodate the volume expansion and stress decreases to -0.18 GPa at the end of sodiation. Upon desodiation, the film initially responds elastically followed by plastic deformation mirroring the response observed during sodiation, but with a peak tensile stress of 0.75 GPa. The stress response of Ge due to sodiation is qualitatively similar to that observed during lithiation; however, the magnitude of stress due to sodiation was significantly less than that due to lithiation. This is interesting, because it is counter intuitive to see that a bigger ion, i.e., sodium, induced lower stress in Ge compared to lithium ion. As a result, the mechanical dissipation losses associated with sodiation are significantly lower compared to lithiation induced dissipative losses. Hence, a better fracture performance and a better cyclic performance can be expected from sodiated Ge than that of lithiated Ge.

**Acknowledgements:** SN would like to acknowledge funding from the National Science Foundation through grant CMMI-1652409 and CMMI-2026717. H.H. would acknowledge the support by National Science Foundation under grant DMR-1742807 and ACS Petroleum Research Fund under grant 58557-ND5.

# References

1. B. L. Ellis and L. F. Nazar, *Curr. Opin. Solid State Mater. Sci.*, **16**, 168–177 (2012).

2. W. Luo et al., *Acc. Chem. Res.*, **49**, 231–240 (2016).

3. E. De La Llave et al., *ACS Appl. Mater. Interfaces*, **8**, 1867–1875 (2016).

4. S. Komaba et al., *Adv. Funct. Mater.*, **21**, 3859–3867 (2011).

5. S. W. Kim, D. H. Seo, X. Ma, G. Ceder, and K. Kang, *Adv. Energy Mater.*, **2**, 710–721 (2012).

6. J.-Y. Hwang, S.-T. Myung, and Y.-K. Sun, *Chem. Soc. Rev.*, **46**, 3529–3614 (2017).

7. K. Kubota and S. Komaba, *J. Electrochem. Soc.*, **162**, A2538–A2550 (2015).

8. H. Moriwake, A. Kuwabara, C. A. J. Fisher, and Y. Ikuhara, *RSC Adv.*, **7**, 36550–36554 (2017).

9. P. Thomas and D. Billaud, *Electrochim. Acta*, **47**, 3303–3307 (2002).

10. C. Bommier and X. Ji, *Isr. J. Chem.*, **55**, 486–507 (2015).

11. L. Baggetto, J. K. Keum, J. F. Browning, and G. M. Veith, *Electrochem. commun.*, **34**, 41–44 (2013).

12. S. Komaba et al., *Electrochem. commun.*, **21**, 65–68 (2012).

13. H. Xie et al., *ACS Energy Lett.*, **3**, 1670–1676 (2018).

14. L. Baggetto et al., *J. Power Sources*, **234**, 48–59 (2013).

15. J. Wang et al., *J. Electrochem. Soc.*, **161**, F3019–F3024 (2014).

16. R. Kali, Y. Krishnan, and A. Mukhopadhyay, *Scr. Mater.*, **130**, 105–109 (2017).

17. P. R. Abel et al., *J. Phys. Chem. C*, **117**, 18885–18890 (2013).

18. M. Dahbi, N. Yabuuchi, K. Kubota, K. Tokiwa, and S. Komaba, *Phys. Chem. Chem. Phys.*, **16**, 15007 (2014).

19. H. Wu and Y. Cui, *Nano Today*, **7**, 414–429 (2012).

20. B. A. Boukamp, *J. Electrochem. Soc.*, **128**, 725 (1981).

21. V. A. Sethuraman, V. Srinivasan, A. F. Bower, and P. R. Guduru, *J. Electrochem. Soc.*, **157**, A1253 (2010).

22. G. Bucci, S. P. V Nadimpalli, V. A. Sethuraman, A. F. Bower, and P. R. Guduru, *J. Mech. Phys. Solids*, **62**, 276–294 (2013).

23. A. Al-Obeidi, D. Kramer, C. V. Thompson, and R. Mönig, *J. Power Sources*, **297**, 472–480 (2015).

24. S. P. V. Nadimpalli, R. Tripuraneni, and V. A. Sethuraman, *J. Electrochem. Soc.*, **162**, A2840–A2846 (2015).

25. M. Pharr, Y. S. Choi, D. Lee, K. H. Oh, and J. J. Vlassak, *J. Power Sources*, **304**, 164–169 (2016).

26. S. K. Soni, B. W. Sheldon, X. Xiao, and A. Tokranov, *Scr. Mater.*, **64**, 307–310 (2011).

27. J. R. Szczech and S. Jin, *Energy Environ. Sci.*, **4**, 56–72 (2011).

28. J. O. Besenhard, J. Yang, and M. Winter, *J. Power Sources*, **68**, 87–90 (1997).

29. L. Y. Beaulieu, K. W. Eberman, R. L. Turner, L. J. Krause, and J. R. Dahn, *Electrochem. Solid-State Lett.*, **4**, A137 (2001).

30. S. P. V. Nadimpalli et al., *J. Electrochem. Soc.*, **160**, A1885–A1893 (2013).

31. V. A. Sethuraman, M. J. Chon, M. Shimshak, V. Srinivasan, and P. R. Guduru, *J. Power Sources*, **195**, 5062–5066 (2010).

32. V. B. Shenoy, P. Johari, and Y. Qi, *J. Power Sources*, **195**, 6825–6830 (2010).

33. H. Yang, W. Liang, X. Guo, C. Wang, and S. Zhang, *Extrem. Mech. Lett.*, **2**, 1–6 (2015).

34. G. Bucci, Y. M. Chiang, and W. C. Carter, *Acta Mater.*, **104**, 33–51 (2016).

35. R. Tripuraneni, S. Rakshit, and S. P. V. Nadimpalli, *J. Electrochem. Soc.*, **165**, A2194–A2202 (2018).

36. A. D. Pelton, *Bull. Alloy Phase Diagrams*, **7**, 25–27 (1986).

37. H. Okamoto, *J. Phase Equilibria Diffus.*, **36**, 644–655 (2015).

38. E. Trofimov, O. Samoilova, O. Zaitseva, and E. Vakhitova, *Metals (Basel).*, **8**, 629 (2018).

39. B. Laforge, L. Levan-Jodin, R. Salot, and A. Billard, *J. Electrochem. Soc.*, **155**, A181 (2008).

40. G. . Stoney, *Proc. R. Soc. London. Ser. A, Contain. Pap. a Math. Phys. Character*, **82**, 172–175 (1909).

41. I. Yoon, D. P. Abraham, B. L. Lucht, A. F. Bower, and P. R. Guduru, *Adv. Energy Mater.*, **6** (2016).

42. L. Y. Beaulieu, A. D. Rutenberg, and J. R. Dahn, *Microsc. Microanal.*, **8**, 422–428 (2002).

43. C. R. Becker, K. E. Strawhecker, Q. P. McAllister, and C. A. Lundgren, *ACS Nano*, **7**, 9173–9182 (2013).

44. V. L. Chevrier and G. Ceder, *J. Electrochem. Soc.*, **158**, A1011 (2011).

45. R. Yazami and Y. Reynier, *J. Power Sources*, **153**, 312–318 (2006).

46. M. H. Tahmasebi, D. Kramer, R. Mönig, and S. T. Boles, *J. Electrochem. Soc.*, **166**, A5001–A5007 (2019).

47. K. J. Rhodes, R. Meisner, M. Kirkham, N. Dudney, and C. Daniel, *J. Electrochem. Soc.*, **159**, A294–A299 (2012).

48. D. H. Nam, K. S. Hong, S. J. Lim, T. H. Kim, and H. S. Kwon, *J. Phys. Chem. C*, **118**, 20086–20093 (2014).

49. J. W. Wang, X. H. Liu, S. X. Mao, and J. Y. Huang, *Nano Lett.*, **12**, 5897–5902 (2012).

50. J. W. Wang et al., *Nano Lett.*, **13**, 709–715 (2013).

51. M. T. McDowell et al., *Nano Lett.*, **13**, 758–764 (2013).

52. E. D. Cubuk and E. Kaxiras, *Nano Lett.*, **14**, 4065–4070 (2014).

53. M. Pharr, Z. Suo, and J. J. Vlassak, *Nano Lett.*, **13**, 5570–5577 (2013).

54. X. Lu et al., *Chem. Mater.*, **28**, 1236–1242 (2016).

55. M. J. Chon, V. A. Sethuraman, A. McCormick, V. Srinivasan, and P. R. Guduru, *Phys. Rev. Lett.*, **107**, 045503 (2011).

56. M. N. Obrovac and L. Christensen, *Electrochem. Solid-State Lett.*, **7**, 93–96 (2004).

57. X. Lu et al., *Chem. Mater.*, **28**, 1236–1242 (2016).